\documentclass[aps,pre,preprint,groupedaddress,showpacs]{revtex4-1}

\usepackage{graphicx}
\usepackage{amssymb}
\usepackage{amsmath}
\usepackage{epstopdf}
\usepackage{natbib}
\usepackage{bm}
\usepackage[utf8]{inputenc}
\usepackage{caption}
\usepackage{float}
\usepackage{subcaption}
\numberwithin{equation}{section}

\def\makeSkob#1#2#3{%
\def\LLL{\mathopen{}\mathclose\bgroup\left} \def\RRR{\aftergroup\egroup\right}
\expandafter \edef \csname #1\endcsname #2##1#3{\SkobInner}
\def\LLL{} \def\RRR{}
\expandafter \edef \csname #1O\endcsname #2##1#3{\SkobInner}
\def\LLL{\bigl} \def\RRR{\bigr}
\expandafter \edef \csname #1A\endcsname #2##1#3{\SkobInner}
\def\LLL{\Bigl} \def\RRR{\Bigr}
\expandafter \edef \csname #1B\endcsname #2##1#3{\SkobInner}
\def\LLL{\biggl} \def\RRR{\biggr}
\expandafter \edef \csname #1C\endcsname #2##1#3{\SkobInner}
\def\LLL{\Biggl} \def\RRR{\Biggr}
\expandafter \edef \csname #1D\endcsname #2##1#3{\SkobInner}
}
\def\SkobInner{\LLL(##1\RRR)} \makeSkob{s}[]
\def\SkobInner{\LLL[##1\RRR]} \makeSkob{sk}[]
\def\SkobInner{\LLL\lbrace##1\RRR\rbrace} \makeSkob{sfig}{}{}
\def\SkobInner{\LLL\lfloor##1\RRR\rfloor} \makeSkob{floor}[]
\def\SkobInner{\LLL\lceil##1\RRR\rceil} \makeSkob{ceil}[]
\def\SkobInner{\LLL\langle##1\RRR\rangle} \makeSkob{ip}<>
\def\SkobInner{\LLL\lvert##1\RRR\rangle} \makeSkob{ket}|>
\def\SkobInner{\LLL\lvert##1\RRR\rvert} \makeSkob{abs}||
\def\SkobInner{\LLL\lVert##1\RRR\rVert} \makeSkob{norm}||
\def\SkobInner{\LLL\lVert##1\RRR\rVert_{\noexpand\mathrm F}} \makeSkob{normFrob}||
\def\SkobInner{\LLL\lVert##1\RRR\rVert_{\noexpand\mathrm{tr}}} \makeSkob{normtr}||

\begin{document}
\title{Gyromagnetic effects in dynamics of magnetic microparticles}
\author{M.Belovs\textsuperscript{1}}
\author{R.Livanovics\textsuperscript{1}}
\author{A.C\={e}bers\textsuperscript{1}}
\email[]{aceb@tesla.sal.lv}

\affiliation{\textsuperscript{1} MMML lab, Department of Physics, University of Latvia, Jelgavas-3, R\={\i}ga, LV-1004, Latvia}

\date{\today}
\begin{abstract}
We derive equations of motion for paramagnetic and ferromagnetic particles fully accounting for gyromagnetic effects. Considering the Einstein-de Haas effect for an ellipsoidal paramagnetic particle we find that starting from a quiescent non-magnetized state, after the field is switched on a rotation along the short axis is established. This is confirmed by the stability analysis of the fixed points of the corresponding ordinary differential equations.
In the case of a ferromagnetic particle we integrate the equations of motion in the dissipationless case by finding the integrals of motion. We also reformulate the equations in a Hamiltonian framework in this case and find a period of small nutation oscillations.
\end{abstract}


\maketitle

\section{Introduction}
The mechanics of levitated objects has recently attracted considerable interest \cite{1}. There are different means of controlling trapped particles among which their control by magnetic fields is of special interest. It has been proposed to use magnetic microparticles as sensors of magnetic fields \cite{2}. The gyromagnetic effects (Einstein-de Haas and Barnett efects) play an important role there. In fact, a new class of interesting mechanical problems arise where in addition to the angular momentum of the body it is necessary to also account for the internal angular momentum due to magnetization \cite{2,3}. Since the particles are levitated in a high vacuum \cite{4} the intrinsic dissipation due to the motion of the magnetic moment with respect to the solid body may play the main role. It may be accounted for by a Landau-Lifshitz type equation for the free particle \cite{5}.

We begin by deriving the coupled set of equations for the particle and its magnetic moment taking into account the internal angular momentum. In part II we consider the paramagnetic particle and on the basis of the derived equations we consider the Einstein-de Haas effect of an axisymmetric ellipsoidal particle which results in the rotation of the particle along its short axis. We then carry out a stability analysis of this regime. Above the critical value of the anisotropy of the magnetic susceptibility the rotating particle enters a precessional regime of motion. In part III we consider the dynamics of a spherical ferromagnetic particle and by finding the integrals of motion in the dissipationless case we show that it is possible to integrate the equations of motion and determine the characteristic frequencies of precession and nutation of the particle. Formulating the Hamiltonian dynamics in this case allows us to apply symplectic algorithms to numerically integrate the equations of the particle and its magnetic moment in the dissipationless case.

\section{Paramagnetic particle}
Magnetic particles used in experiments may be in a multidomain state. We may take this into account by considering the dynamics of a paramagnetic microparticle.

\subsection*{Model}
The energy of an axisymmetric ellipsoidal paramagnetic particle taking into account the effect of the demagnetizing field and the rotational kinetic energy reads
\begin{equation}
E=\frac{\vec{m}^{2}}{2\chi}+\frac{1}{2}N_{\perp}\vec{m}_{\perp}^{2}+\frac{1}{2}N_{\parallel}\vec{m}^{2}_{\parallel}-\vec{m}\cdot\vec{H}+\frac{\vec{\Omega}\cdot\vec{L}}{2}~,
\label{Eq:1}
\end{equation}
where $\vec{m}$ is the magnetic moment of the particle, $\vec{m}_{\parallel},\vec{m}_{\perp}$ denote its components along the symmetry axis and perpendicularly to it respectively, $\vec{L}$ is the angular momentum of the particle and $\vec{\Omega}$ is its angular velocity. The relation (\ref{Eq:1}) may be rewritten as follows
\begin{equation}
E=\frac{\vec{m}^{2}}{2\chi_{0}}-\frac{1}{2}(N_{\perp}-N_{\parallel})\vec{m}^{2}_{\parallel}-\vec{m}\cdot\vec{H}+\frac{1}{2}\vec{\Omega}\cdot\vec{L}~,
\label{Eq:2}
\end{equation}
where $\chi_{0}=\chi/(1+\chi N_{\perp})$, which in the case of a spherical particle when $N_{\perp}=4\pi/3$ is its magnetic susceptibility taking into account the effect of the demagnetizing field.

We introduce the unit vectors $\vec{e}_{i}$ ($i=1,2,3$) along the principal axes of inertia ($\vec{e}_{3}$ is along the symmetry axis of an axisymmetric prolate ellipsoid). As a result the energy may written as follows
\begin{equation}
E=\frac{\vec{m}^{2}}{2\chi_{0}}-\frac{1}{2}(N_{\perp}-N_{\parallel})(\vec{e}_{3}\cdot\vec{m})^{2}-\vec{m}\cdot\vec{H}+\frac{1}{2}\vec{\Omega}\cdot\vec{L}~.
\label{Eq:2a}
\end{equation}
We calculate $\frac{dE}{dt}$ taking into account the equation for the total angular momentum ($\gamma$ is the gyromagnetic ratio)
\begin{equation}
\frac{d\vec{L}}{dt}+\frac{1}{\gamma}\frac{d\vec{m}}{dt}=[\vec{m}\times\vec{H}]
\label{Eq:3}
\end{equation}
and
\begin{equation}
\frac{d\vec{e}_{3}}{dt}=[\vec{\Omega}\times\vec{e}_{3}]
\label{Eq:4}
\end{equation}
and as a result we have
\begin{equation}
\frac{dE}{dt}=\Bigl(\frac{\vec{m}}{\chi_{0}}-\vec{H}-\vec{H}_{a}-\frac{\vec{\Omega}}{\gamma}\Bigr)\frac{d\vec{m}}{dt}+\vec{\Omega}\cdot[\vec{m}\times(\vec{H}+\vec{H}_{a})]~,
\label{Eq:5}
\end{equation}
where $\vec{H}_{a}=(N_{\perp}-N_{\parallel})(\vec{e}_{3}\cdot\vec{m})\vec{e}_{3}$.
$\frac{d\vec{m}}{dt}$ contains non-dissipative and dissipative terms. Relation (\ref{Eq:5}) may be rewritten as follows
\begin{equation}
\frac{dE}{dt}=\Bigl(\frac{\vec{m}}{\chi_{0}}-\vec{H}-\vec{H}_{a}-\frac{\vec{\Omega}}{\gamma}\Bigr)\Bigl(\frac{d\vec{m}}{dt}-[\vec{\Omega}\times\vec{m}]\Bigr)
\label{Eq:6}
\end{equation}
and we see that the non-dissipative contribution may be put as follows
\begin{equation}
\Bigl(\frac{d\vec{m}}{dt}-\vec{\Omega}\times\vec{m}]\Bigr)=-\gamma\Bigl[\Bigl(\vec{H}+\vec{H}_{a}+\frac{\vec{\Omega}}{\gamma}\Bigr)\times\vec{m}\Bigr]~.
\label{Eq:7}
\end{equation}
For the dissipative part we take a linear phenomenological relation
\begin{equation}
\Bigl(\frac{d\vec{m}}{dt}-[\vec{\Omega}\times\vec{m}]\Bigr)_{diss}=-\frac{\chi_{0}}{\tau}\Bigl(\frac{\vec{m}}{\chi_{0}}-\vec{H}-\vec{H}_{a}-\frac{\vec{\Omega}}{\gamma}\Bigr)~.
\label{Eq:8}
\end{equation}
As a result we have an equation for the magnetic moment of the particle that takes into account its gyromagnetic properties
\begin{equation}
\frac{d\vec{m}}{dt}=\gamma[\vec{m}\times(\vec{H}+\vec{H}_{a})]-\frac{1}{\tau}\Bigl(\vec{m}-\chi_{0}\Bigl(\vec{H}+\vec{H}_{a}+\frac{\vec{\Omega}}{\gamma}\Bigr)\Bigr)~.
\label{Eq:9}
\end{equation}
In the case $\vec{H}_{a}=0$ Eq.(\ref{Eq:9}) was considered for a paramagnetic fluid with gyromagnetic properties in \cite{6} and some interesting hydrodynamic phenomena were predicted, in particular for liquid oxygen.

\subsection*{Einstein-de Haas effect for ellipsoidal paramagnetic particle}
Here we address the question of how gyromagnetism affects the behavior of a paramagnetic particle. For example, if we have an anisotropic non-rotating, non-magnetized paramagnetic particle and switch on a magnetic field the particle will start to rotate. The question is about which axis this will take place.
We resolve this question by numerical simulation of the full set of equations. In the case $\vec{H}_{a}=0$ the
equation for the magnetic moment in the body frame of reference reads ($d'/dt$ denotes the derivative with respect to the body frame)
\begin{equation}
\frac{d'\vec{m}}{dt}=\gamma[\vec{m}\times\Bigl(\vec{H}+\frac{\vec{\Omega}}{\gamma}\Bigr)]-\frac{1}{\tau}\Bigl(\vec{m}-\chi_{0}\Bigl(\vec{H}+\frac{\vec{\Omega}}{\gamma}\Bigr)\Bigr)~.
\label{Eq:10}
\end{equation}
The equation for the mechanical angular momentum we obtain from Eq.(\ref{Eq:3}) and in the body frame it reads
\begin{equation}
\frac{d'\vec{L}}{dt}+[\vec{\Omega}\times\vec{L}]=\frac{1}{\tau\gamma}\Bigl(\vec{m}-\chi_{0}\Bigl(\vec{H}+\frac{\vec{\Omega}}{\gamma}\Bigr)\Bigr)~,
\label{Eq:11}
\end{equation}
where for an axisymmetric particle $\vec{L}=I_{\perp}(\Omega_{1}\vec{e}_{1}+\Omega_{2}\vec{e}_{2})+I_{\parallel}\Omega_{3}\vec{e}_{3}$.

\begin{figure}
	\centering
	\includegraphics[width=0.9\textwidth]{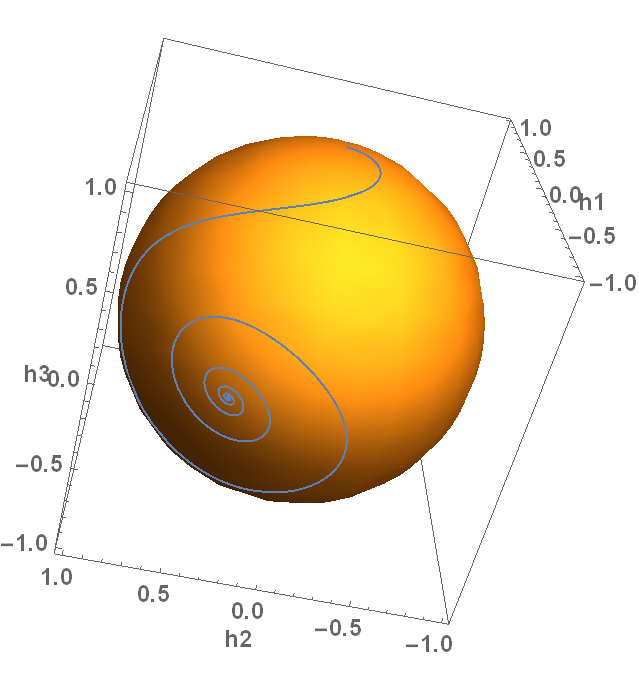}
	\caption{Einstein-de Haas effect of ellipsoidal paramagnetic particle. Long axis relaxes perpendicularly to applied field. $\omega_{0}=0.6;~\sigma=5;~\delta=1.5$. }
	\label{fig:1}
\end{figure}

The following scalings are introduced $\vec{\Omega}=|\gamma|H\tilde{\vec{\Omega}}~ (\gamma<0);~\vec{m}=\chi_{0}H\tilde{\vec{m}};~\vec{L}=I_{\parallel}|\gamma|H\tilde{\vec{L}};~t=\tau\tilde{t}$. Taking as the parameters $\omega_{0}=|\gamma|H\tau;~\delta=\frac{\chi_{0}}{|\gamma|^{2}I_{\parallel}};~\sigma=\frac{I_{\perp}}{I_{\parallel}}$ we have the set of dimensionless equations (tildas are omitted).
\begin{equation}
\frac{d'\vec{m}}{dt}=-\omega_{0}[\vec{m}\times(\vec{h}-\vec{\Omega})]-(\vec{m}-\vec{h}+\vec{\Omega})~,
\label{Eq:12}
\end{equation}
\begin{equation}
\frac{d'\vec{L}}{dt}+\omega_{0}[\vec{\Omega}\times\vec{L}]=-\delta(\vec{m}-\vec{h}+\vec{\Omega})~.
\label{Eq:13}
\end{equation}
We consider the case when the applied field is constant after switching on. Thus, in the body frame for the field $\vec{h}$ we have the equation
\begin{equation}
\frac{d'\vec{h}}{dt}=-\omega_{0}[\vec{\Omega}\times\vec{h}]~.
\label{Eq:14}
\end{equation}
To find the orientation of the particle in the laboratory frame we add the equations for the unit vectors along the principal axes
\begin{equation}
\frac{d\vec{e}_{i}}{dt}=\omega_{0}[\vec{\Omega}\times\vec{e}_{i}]~(i=1,2,3)~.
\label{Eq:15}
\end{equation}
As a result we have a set of 18 equations which are solved with corresponding initial conditions. Particularly interesting is the Einstein-de Haas effect for an anisotropic particle. In this case we take the following initial conditions
\begin{eqnarray}
\vec{\Omega}(0)=0;~\vec{m}(0)=0; \vec{e}_{1}(0)=(\cos{(\vartheta)},0,\sin{(\vartheta)});~\vec{e}_{2}(0)=(0,1,0);\\ \nonumber~\vec{e}_{3}=(-\sin{(\vartheta)},0,\cos{(\vartheta)});
 h_{1}(0)=\sin{(\vartheta)},h_{2}(0)=0;~h_{3}(0)=\cos{(\vartheta)}~.
 \label{Eq:17}
\end{eqnarray}
The trajectory on the unit sphere of the unit vector $\vec{h}$ in the body frame at initial conditions (II.17) with $\vartheta=0.1$ and the parameters $\omega_{0}=0.6;~\sigma=5;~\delta=1.5$ is shown in Fig.\ref{fig:1}.
\begin{figure}
	\centering
	\includegraphics[width=0.9\textwidth]{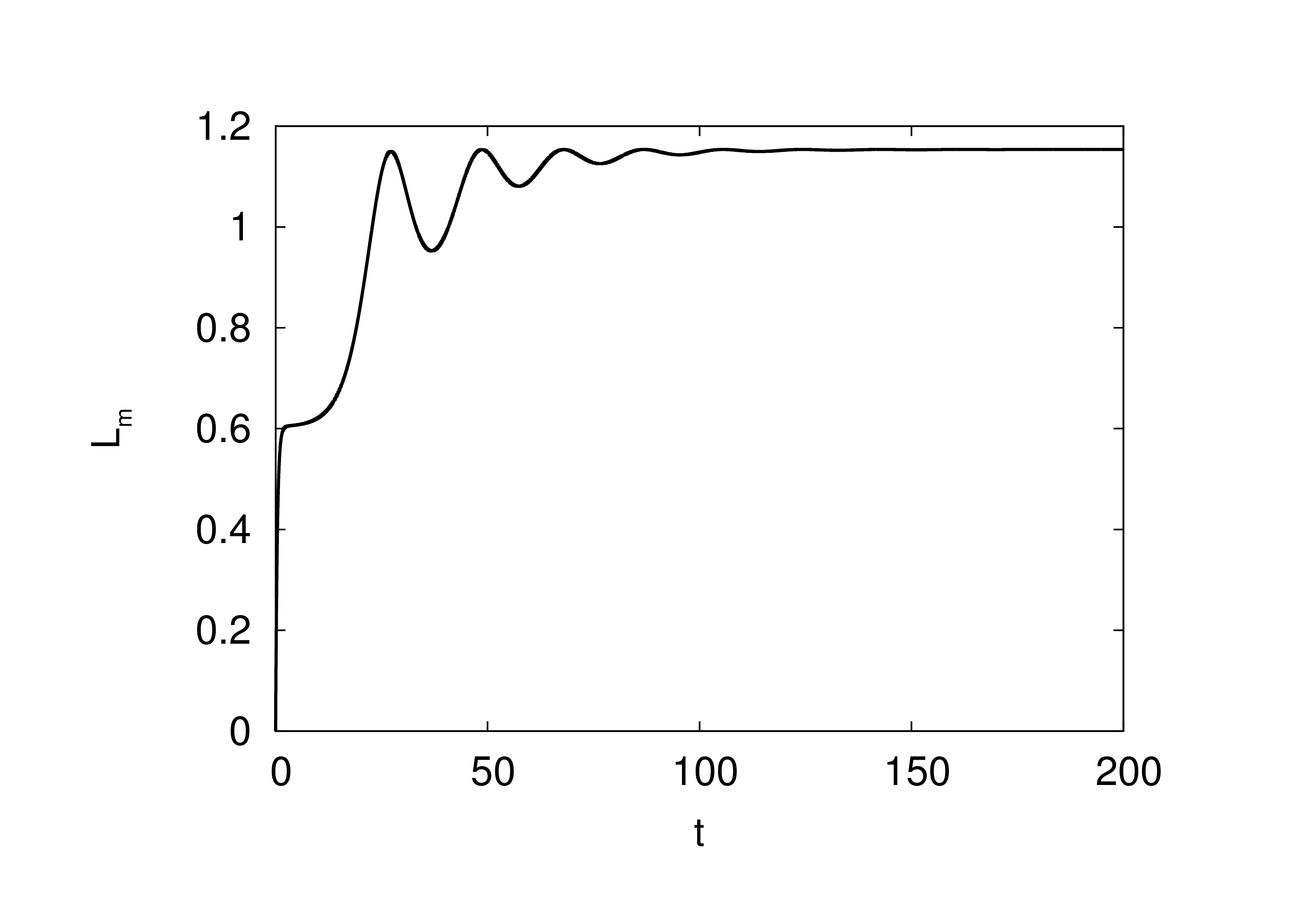}
	\caption{Mechanical angular momentum as a function of time. $\omega_{0}=0.6;~\sigma=5;~\delta=1.5$. }
	\label{fig:2a}
\end{figure}
\begin{figure}
	\centering
	\includegraphics[width=0.9\textwidth]{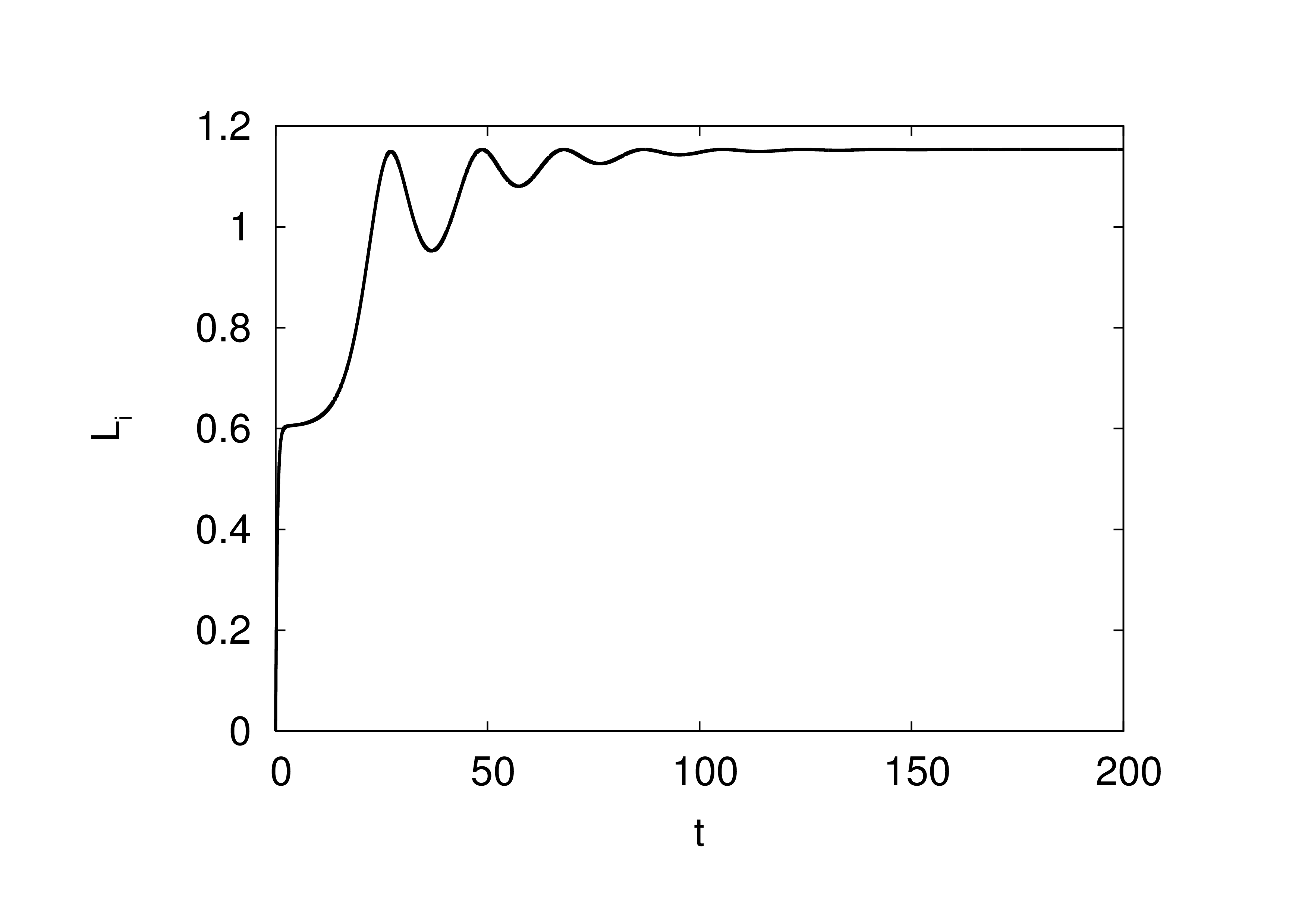}
	\caption{Internal angular momentum as a function of time. $\omega_{0}=0.6;~\sigma=5;~\delta=1.5$. }
	\label{fig:2b}
\end{figure}
We see that a particle with a long axis orientation close to the applied field relaxes to a state where the long axis orients itself perpendicularly to the applied field. Thus the Einstein-de Haas effect for the anisotropic particle results in the rotation of the particle around some short axis, the orientation of which in $\vec{e}_{1},\vec{e}_{2}$ plane depends on the initial conditions. Mechanical $L_{m}=\sigma\Omega_{1}e_{1z}+\sigma\Omega_{2}e_{2z}+\Omega_{3}e_{3z}$ and internal $L_{i}=\delta(m_{1}e_{1z}+m_{2}e_{2z}+m_{3}e_{3z})$ angular moments along the field direction are shown in Fig.\ref{fig:2a},\ref{fig:2b}. We  see that the $z$ component of the total angular momentum $L_{m}-L_{i}$ is zero, as expected due to the conservation of the total angular  momentum component along the field direction. The observed Einstein-de Haas effect poses a question about the stability of the rotation of the particle about its long and short axes. This is studied by a linear stability analysis.

The relaxation of the particle to the state where the long axis is perpendicular to the applied field may be compared with the well-known demonstration of the spinning egg \cite{7}, when a hard-boiled egg spinning on a hard surface with an initially horizontal axis of symmetry gradually rises until the axis of symmetry is fully vertical. In this case the mechanism of dissipation at sliding contact is important and causes the relaxation to the minimal energy state \cite{8}. In our case there is an internal mechanism of dissipation caused by the magnetic relaxation taking place even as the angular momentum is conserved.
Let us compare the energies for the cases with the long axis parallel and perpendicular to the field (in dimensional units). In the case of rotation along the long axis $\vec{\Omega}=(0,0,\Omega_{3})$; $\vec{m}=\chi_{0}(0,0,H_{3}+\Omega_{3}/\gamma)$ the energy reads (neglecting magnetic anisotropy)
\begin{equation}
E=\frac{\vec{m}^{2}}{2\chi_{0}}-\vec{m}\cdot\vec{H}+\frac{1}{2}\vec{\Omega}\cdot\vec{L}=\frac{\chi_{0}\Omega^{2}_{3}}{2\gamma^{2}}-\frac{\chi_{0}H^{2}_{3}}{2}+\frac{1}{2}\Omega^{2}_{3}I_{\parallel}~.
\label{Eq:20a}
\end{equation}
Taking into account conservation of the total angular momentum (in the initial state we have $\vec{\Omega}=0;~\vec{m}=0$) and $I_{\parallel}\Omega_{3}+m_{3}/\gamma=0$ which yields $\Omega_{3}=-\frac{1}{2\gamma}\frac{\chi_{0}H_{3}}{I_{\parallel}+\chi_{0}/\gamma^{2}}$. Inserting $\Omega_{3}$ in relation (\ref{Eq:20a}) we obtain ($H_{3}=H$)
\begin{equation}
E_{l}=-\frac{1}{2}\frac{I_{\parallel}H^{2}}{I_{\parallel}+\chi_{0}/\gamma^{2}}~.
\label{Eq:21a}
\end{equation}
In the case of rotation along the short axis we similarly obtain
\begin{equation}
E_{s}=-\frac{1}{2}\frac{I_{\perp}H^{2}}{I_{\perp}+\chi_{0}/\gamma^{2}}~.
\label{Eq:21b}
\end{equation}
Since $I_{\perp}>I_{\parallel}$ we see that $E_{s}<E_{l}$ and thus in the case of the rotation along the short axis the energy is smaller and we have the analogy with the case of the spinning egg.

\subsection*{Stability}
Let us consider the stability of the fixed point of the set of differential equations (\ref{Eq:12},\ref{Eq:13},\ref{Eq:14}) $\vec{\Omega}=(0,0,\Omega^{0}_{3});~\vec{m}=(0,0,1-\Omega^{0}_{3}),~\vec{h}=(0,0,1)$. Its perturbation is introduced as follows (perturbations of $\Omega_{3}$ and $m_{3}$ do not influence stability) $\vec{\Omega}=(\omega_{1},\omega_{2},\Omega^{0}_{3});~\vec{m}=(\mu_{1},\mu_{2},1-\Omega^{0}_{3});~\vec{h}=(s_{1},s_{2},1)$ and introducing $\omega=\omega_{1}+I\omega_{2};~\mu=\mu_{1}+I\mu_{2};~s_{1}+Is_{2}$ for small perturbations we obtain an equation for the vector $w=(\omega,\mu,s)^{T}$ 
\begin{eqnarray}
\sigma\frac{d\omega}{dt}=-I\omega_{0}(\sigma-1)\Omega^{0}_{3}\omega-\delta(\mu-s+\omega)\\ \nonumber
\frac{d\mu}{dt}=(1-I\omega_{0}m^{0}_{3})(s-\mu-\omega)\\ \nonumber
\frac{ds}{dt}=I\omega_{0}\omega-I\Omega^{0}_{3}\omega_{0}s~.
\label{Eq:18}
\end{eqnarray}
\begin{figure}
	\centering
	\includegraphics[width=0.9\textwidth]{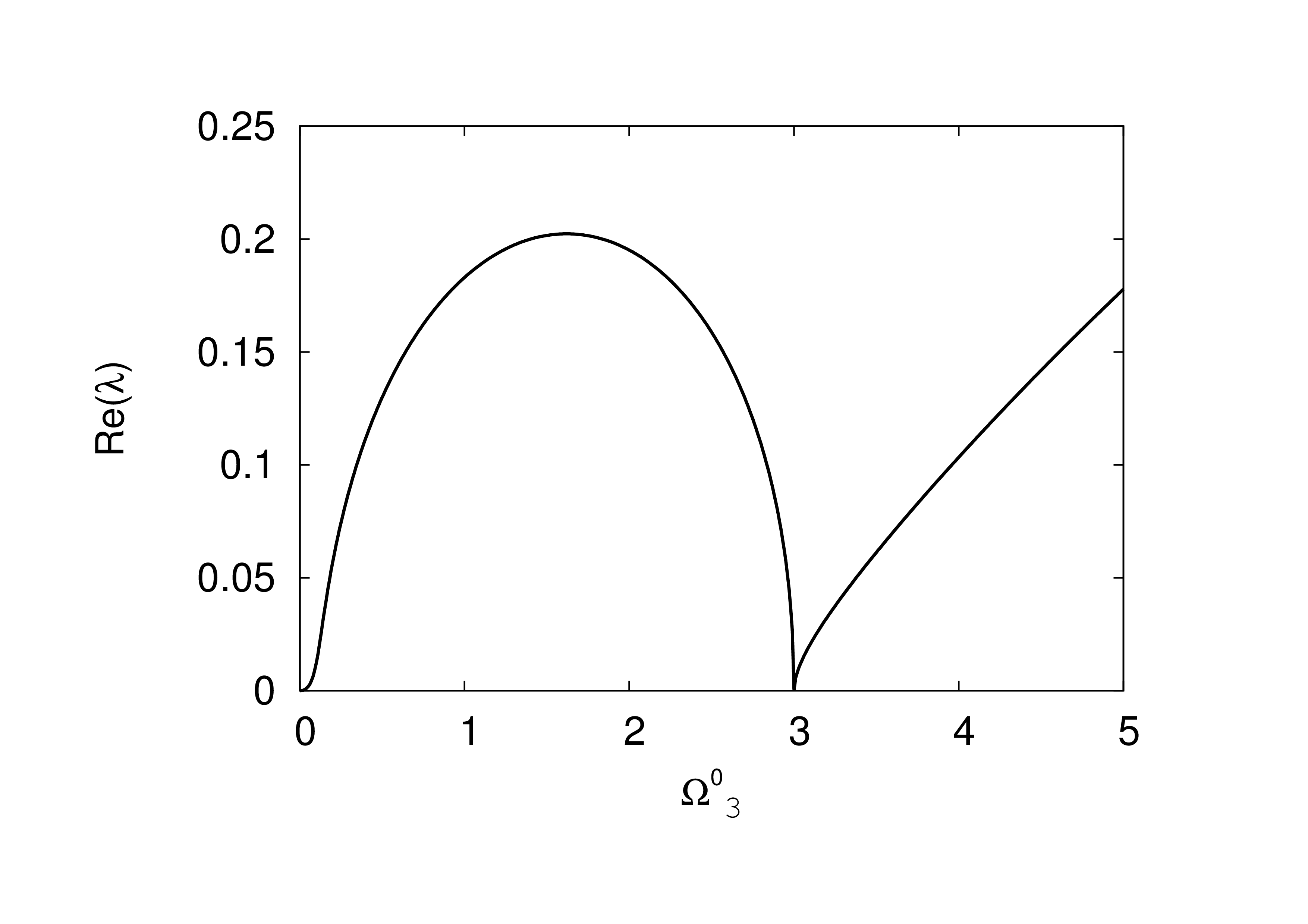}
	\caption{Largest real part of eigenvalues as a function of $\Omega^{0}_{3}$. $\omega_{0}=0.6;~\sigma=3;~\delta=1.5$. }
	\label{fig:3}
\end{figure}
Calculation of the eigenvalues $\lambda$ of the matrix $A$ ($\frac{dw}{dt}=Aw$) shows that they are complex and one of them has a positive real part. $max(Re(\lambda))$ in dependence on $\Omega^{0}_{3}$ as a parameter is shown in Fig.\ref{fig:3} ($\omega_{0}=0.6;~\delta=1.5;~\sigma=3$). It is interesting to note the case $\Omega^{0}_{3}=\sigma$ when this eigenvalue is imaginary.
\begin{figure}
	\centering
	\includegraphics[width=0.9\textwidth]{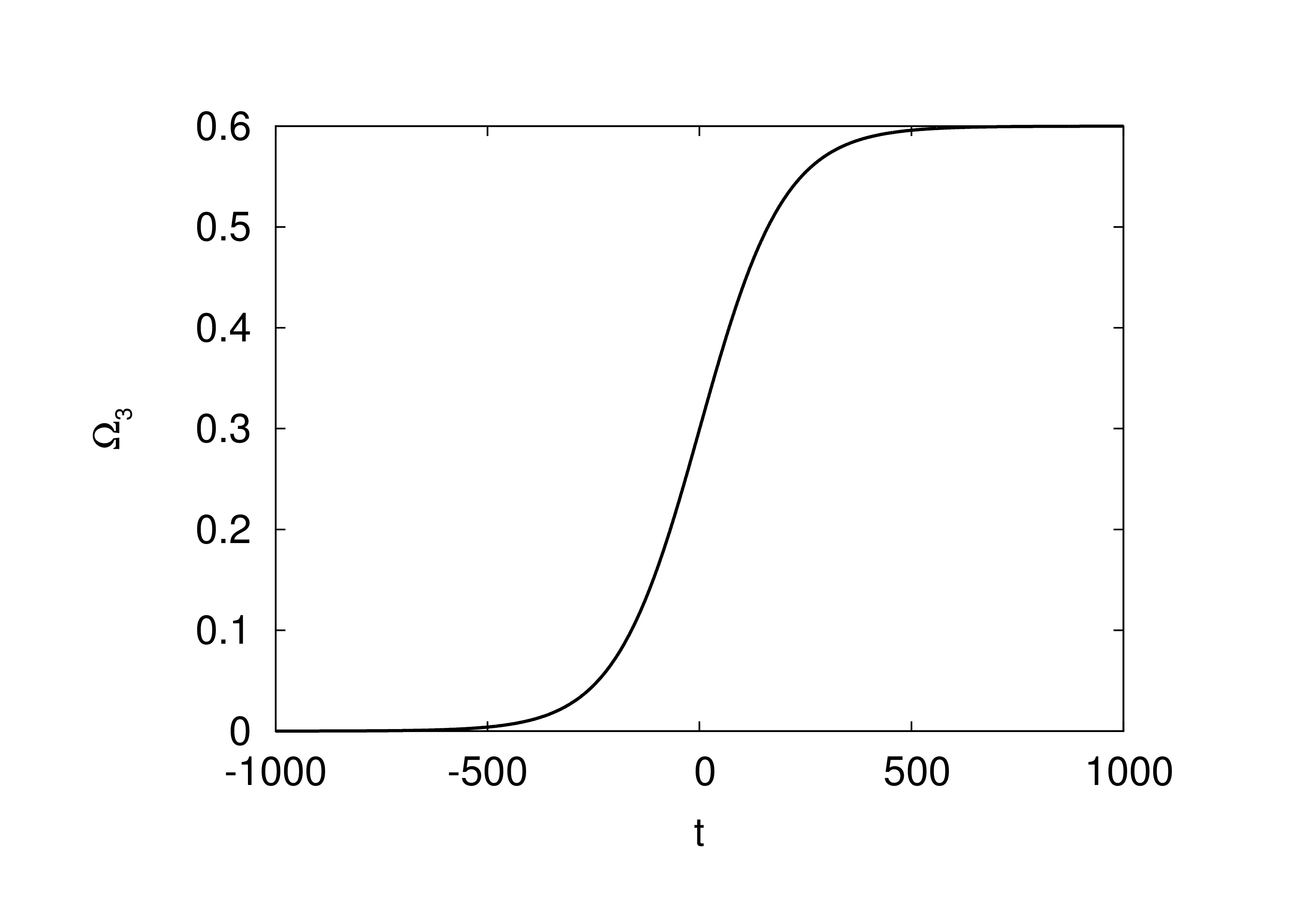}
	\caption{Angular velocity at adiabatic switching of the magnetic field, $\Omega^{0}_{3}$. $\omega_{0}=0.6;~\sigma=5;~\delta=1.5;~\varepsilon=0.01$.}
	\label{fig:3c}
\end{figure}

We should note that the rotation along the long axis may be realized at adiabatic switching of the magnetic field when dissipation is absent and the instability mechanism considered above does not work. Taking for the field $H=f(t)$ where $f(t)=1/(1+\exp{(-\varepsilon t)})$ and starting from the initial conditions $\vec{\Omega}=0;~\vec{m}=0$ we achieve the steady state with (in dimensionless units) $\Omega_{3}=\delta/(1+\delta)$ and $m_{3}=1/(1+\delta)$ as found from the conservation of total angular momentum and the equilibrium value of the magnetization $\Omega_{3}=\delta m_{3}$ and $m_{3}=h_{3}-\Omega_{3}$. This is shown in \ref{fig:3c} for $\omega_{0}=0.6;~\sigma=5;~\delta=1.5;~\varepsilon=0.01$.
 
More complex is the study of the stability of rotation along the short axis. Perturbations of the stationary state $\Omega_{1}^{0};~\Omega^{0}_{2}=0;~\Omega^{0}_{3}=0;~\vec{m}^{0}=(h^{0}_{1}-\Omega^{0}_{1},0,0);~\vec{h}=)1,0,0)$ are introduced as follows $\vec{\Omega}=(\Omega^{0}_{1},\omega_{2},\omega_{3});~\vec{m}=(h^{0}_{1}-\Omega^{0}_{1},\mu_{2},\mu_{3});~\vec{h}=(1,s_{2},s_{3})$. The linear set of equations for small perturbations reads
\begin{eqnarray}
\frac{d\omega_{2}}{dt}=\omega_{0}\Omega^{0}_{1}(\frac{1}{\sigma}-1)\omega_{3}-\frac{\delta}{\sigma}(\mu_{2}-s_{2}+\omega_{2})\\ \nonumber
\frac{d\omega_{3}}{dt}=-\delta(\mu_{3}-s_{3}+\omega_{3})\\ \nonumber
\frac{d\mu_{2}}{dt}=\omega_{0}(h^{0}_{1}-\Omega^{0}_{1})(s_{3}-\mu_{3}-\omega_{3})-(\mu_{2}-s_{2}+\omega_{2})\\ \nonumber
\frac{d\mu_{3}}{dt}=-\omega_{0}(h^{0}_{1}-\Omega^{0}_{1})(s_{2}-\mu_{2}-\omega_{2})-(\mu_{3}-s_{3}+\omega_{3})\\ \nonumber
\frac{ds_{2}}{dt}=-\omega_{0}(\omega_{3}h^{0}_{1}-\Omega^{0}_{1}s_{3}) \\ \nonumber
\frac{ds_{3}}{dt}=-\omega_{0}(\Omega^{0}_{1}s_{2}-\omega_{2}h^{0}_{1})\nonumber ~.
\label{Eq:19}
\end{eqnarray}
\begin{figure}
	\centering
	\includegraphics[width=0.9\textwidth]{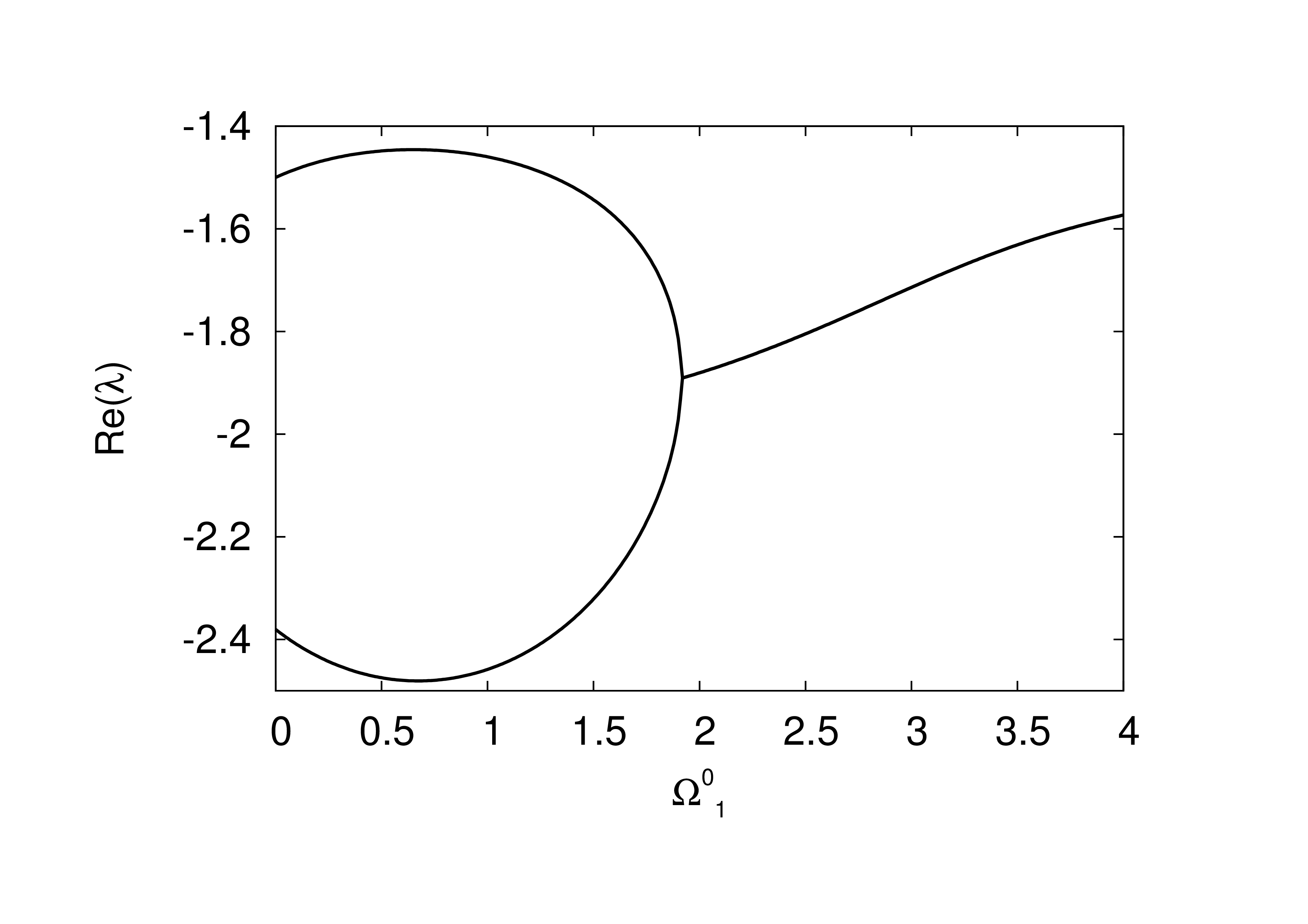}
	\caption{Real parts of a pair of complex conjugated eigenvalues as a function of $\Omega^{0}_{1}$. $\omega_{0}=0.6;~\sigma=3;~\delta=1.5$. }
	\label{fig:4a}
\end{figure}
\begin{figure}
	\centering
	\includegraphics[width=0.9\textwidth]{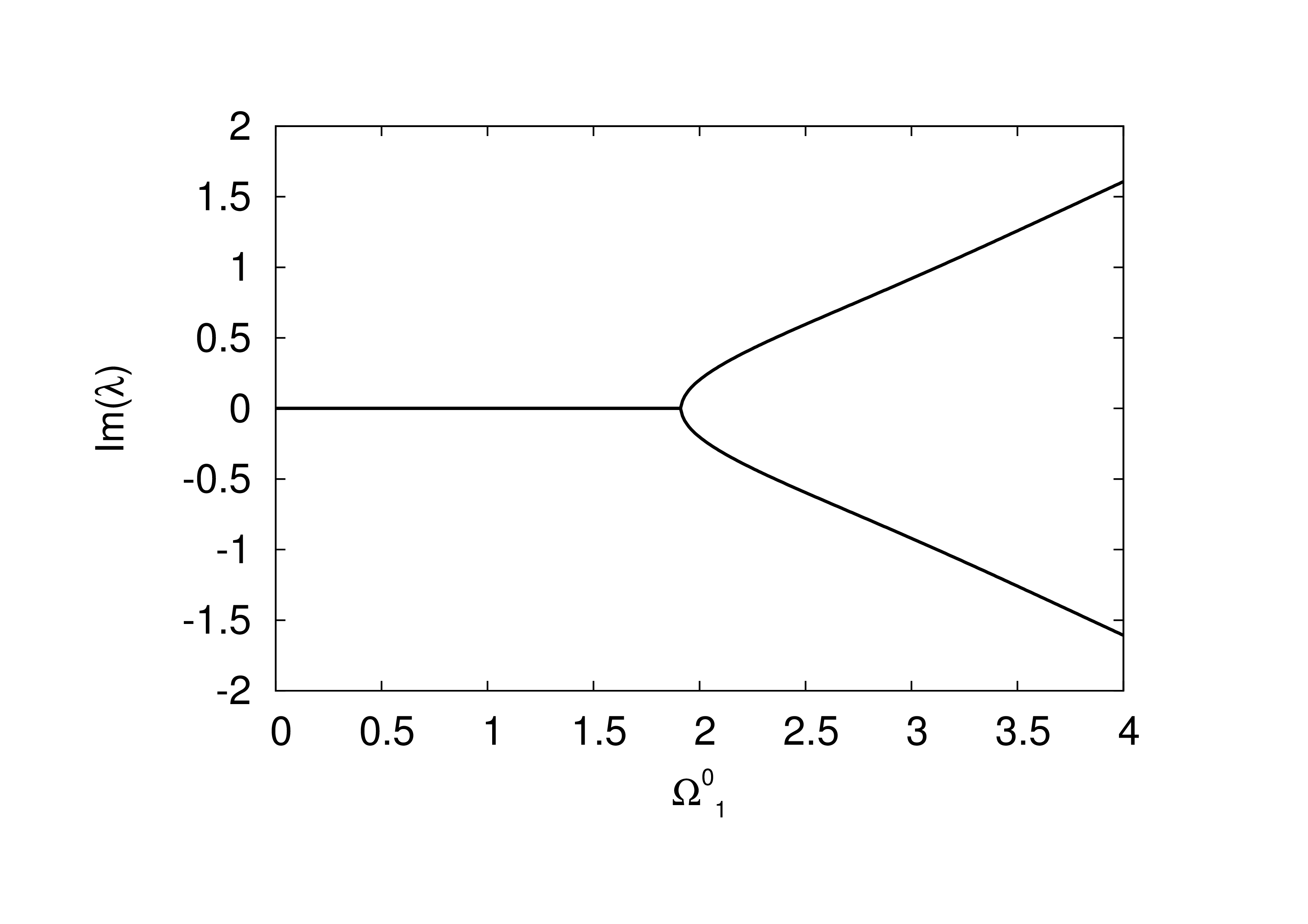}
	\caption{Imaginary parts of a pair of complex conjugated eigenvalues as a function of $\Omega^{0}_{1}$. $\omega_{0}=0.6;~\sigma=3;~\delta=1.5$. }
	\label{fig:4b}
\end{figure}
Two pairs of eigenvalues of the matrix of the set (II.19) are complex conjugated with negative real parts, two are real, of which one is zero and other negative. Real and imaginary parts of one pair are shown in Fig.\ref{fig:4a},\ref{fig:4b}. It is interesting to note the double degeneracy of the zero eigenvalue at $\Omega^{0}_{1}=1$. As a result it is shown that the rotation along the short axis is stable as was already found when considering the Einstein-de Haas effect.

\subsection*{Stationary solution}
In the dissipationless case it is possible to find stationary solutions with the angular velocity of the particle describing the trajectory on the surface of an ellipsoid. In the stationary case the magnetic moment is constant in the body frame. Since ($\vec{z}=\vec{h}-\vec{m}-\vec{\Omega}$)
\begin{equation}
\frac{d'\vec{m}}{dt}=-\omega_{0}[\vec{m}\times\vec{z}]+\vec{z}=B\vec{z}
\label{Eq:20}
\end{equation}
and $\det(B)=1+\omega_{0}^{2}(m_{1}^{2}+m_{2}^{2}+m_{3}^{2})>0$~,
then the only possibility is that $\vec{z}=0$. Since $\Omega_{3}=\Omega_{3}^{0}$ is the integral of motion then $h_{3}=\cos{(\vartheta)}=const$ and we have ($\gamma_{0}=\omega_{0}(1-\frac{1}{\sigma});~\omega=\gamma_{0}\Omega^{0}_{3}$)
\begin{equation}
\frac{d\Omega_{1}}{dt}=\gamma_{0}\Omega_{3}^{0}\Omega_{2};~\frac{d\Omega_{2}}{dt}=-\gamma_{0}\Omega_{3}^{0}\Omega_{1}~.
\label{Eq:21}
\end{equation}
We consider the Euler angles $\varphi,\vartheta,\psi$ \cite{9}. Fixing the phase of the solution of Eqs.(\ref{Eq:21}) we have
\begin{equation}
\Omega_{1}=\Omega_{\perp}\sin{(\psi)};~\Omega_{2}=\Omega_{\perp}\cos{(\psi)};~\psi=\frac{\pi}{2}-\varphi_{0}+\omega t ~.
\label{Eq:22}
\end{equation}
Since $h_{3}=\cos{(\vartheta)}=const$ then $\Omega_{1}h_{2}=\Omega_{2}h_{1}$ and as a result $h_{1}=\sin{(\vartheta)}\sin{(\psi)};~h_{2}=\sin{(\vartheta)}\cos{(\psi)}$. The condition of stationarity of $m_{1},m_{2}$ gives $\Omega_{\perp}=\sin{(\vartheta)}$.
Relations for the angular velocity in the body frame \cite{7} ($\vartheta=const$)
\begin{equation}
\omega_{0}\Omega_{1}=\dot{\varphi}\sin{(\vartheta)}\sin{(\psi)};~\omega_{0}\Omega_{3}=\dot{\varphi}\sin{(\vartheta)}\cos{(\psi)};~\omega_{0}\Omega_{3}=\dot{\varphi}\cos{(\vartheta)}+\dot{\psi}~.
\label{Eq:23}
\end{equation}
As a result we obtain $\dot{\varphi}=\omega_{0}$ and $\Omega_{3}=\sigma\cos{(\vartheta)}$. Thus in the stationary case the angular velocity is on surface of the ellipsoid
\begin{equation}
\frac{\Omega^{2}_{3}}{\sigma^{2}}+\Omega^{2}_{1}+\Omega^{2}_{2}=1~.
\label{Eq:24}
\end{equation}

It is interesting to consider the stability of the dissipationless solution. Dynamics with the dissipation taken into account starting from initial state
\begin{eqnarray}
e_{1x}=0;e_{1y}=\cos{(\vartheta)};e_{1z}=\sin{(\vartheta)},~e_{2x}=-1;e_{2y}=0;e_{2z}=0,\\ \nonumber
e_{3x}=0;e_{3y}=-\sin{(\vartheta)};e_{3z}=\cos{(\vartheta)},h_{1}=\sin{(\vartheta)};h_{2}=0;h_{3}=\cos{(\vartheta)},\\ \nonumber
\Omega_{1}=\sin{(\vartheta)};\Omega_{2}=0;\Omega_{3}=\sigma\cos{(\vartheta)}~.
\label{Eq:25}
\end{eqnarray}
\begin{figure}
	\centering
	\includegraphics[width=0.9\textwidth]{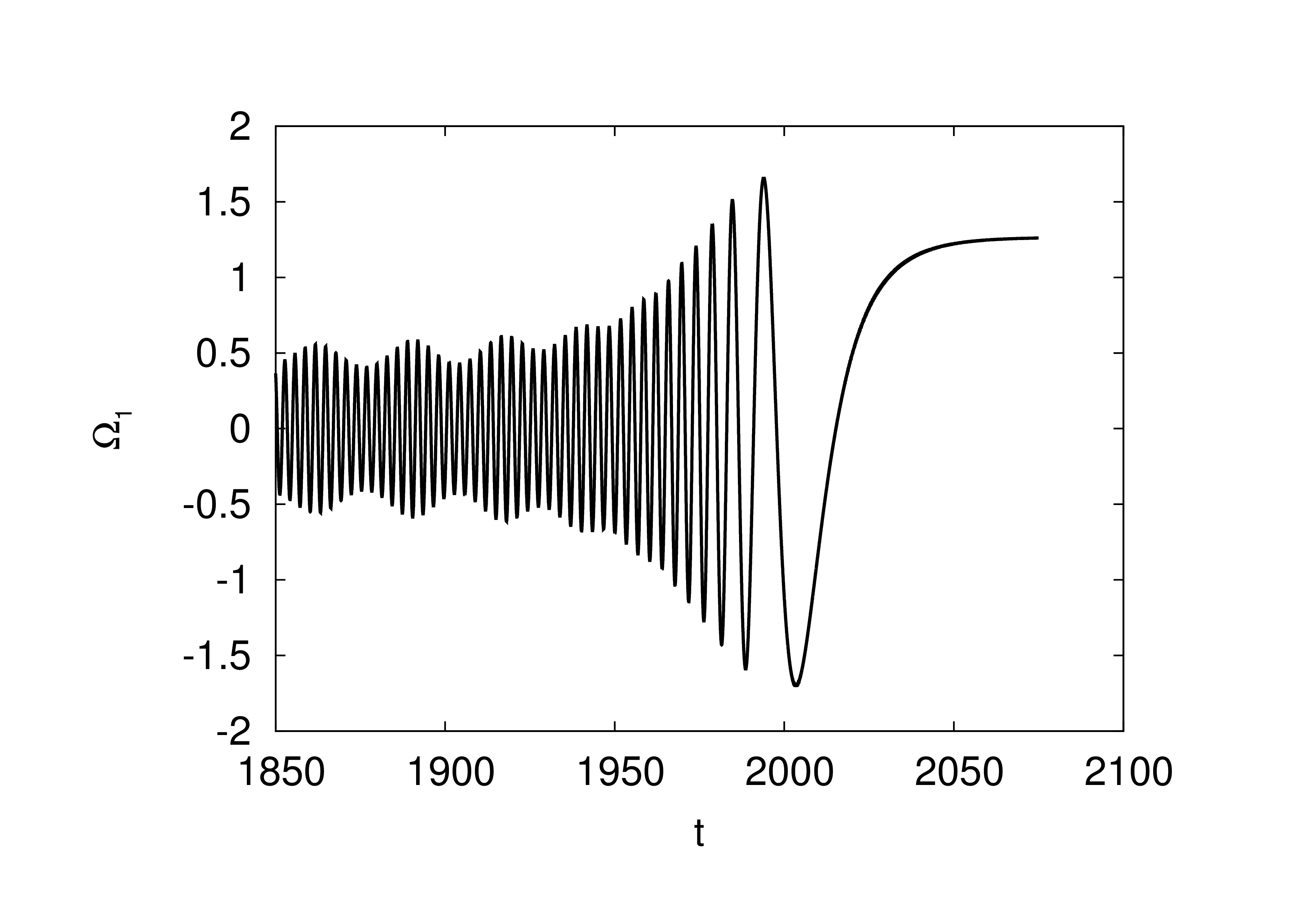}
	\caption{Relaxation of stationary solution to rotation along short axis. $\omega_{0}=0.6;~\sigma=5;~\delta=1.5$. Initial condition (\ref{Eq:25}) at $\vartheta(0)=0.5$. }
	\label{fig:5}
\end{figure}
\begin{figure}
	\centering
	\includegraphics[width=0.9\textwidth]{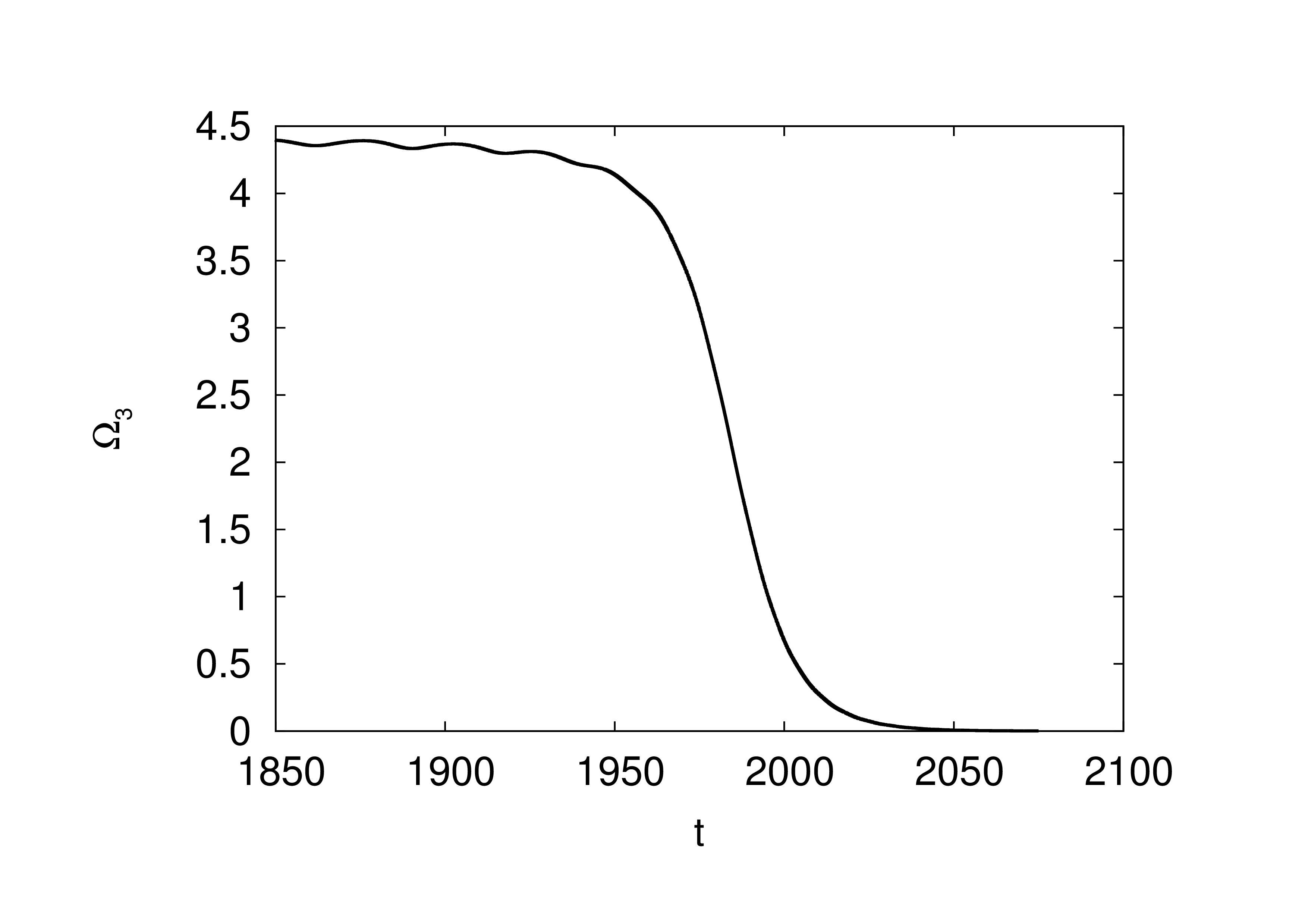}
	\caption{Relaxation of angular velocity along long axis of particle.$\omega_{0}=0.6;~\sigma=5;~\delta=1.5$. Initial condition (\ref{Eq:25}) at $\vartheta=0.5$. }
	\label{fig:5b}
\end{figure}
is shown in Fig.\ref{fig:5} and Fig.\ref{fig:5b}. We see that the system moves along the dissipationless trajectory for a quite long time until it finally exhibits growth of perturbations and then relaxes to a state with the angular velocity along the short axis. In the final state the long axis is perpendicular to the field and $\Omega_{3}=0$. To have this instability the dissipation is necessary.

\subsection*{Paramagnetic particle with anisotropy of magnetic susceptibility}
Let us consider the dynamics of a particle with torque due $\vec{H_{a}}$ taken into account. Using the scalings as given above we have the following set of equations in the body frame (tildas are omitted)
\begin{equation}
\frac{d'\vec{m}}{dt}=-\omega_{0}[\vec{m}\times(\vec{h}+\vec{h}_{a}-\vec{\Omega})]-(\vec{m}-\vec{h}-\vec{h}_{a}+\vec{\Omega})
\label{Eq:26}
\end{equation}
and
\begin{equation}
\frac{d'\vec{L}}{dt}+\omega_{0}[\vec{\Omega}\times \vec{L}]=-\delta\omega_{0}[\vec{m}\times\vec{h}_{a}]-\delta(\vec{m}-\vec{h}-\vec{h}_{a}+\vec{\Omega})~,
\label{Eq:27}
\end{equation}
where $\vec{h}_{a}=k\vec{e}_{3}(\vec{e}_{3}\cdot\vec{m})$ $(k=\chi_{0}(N_{\perp}-N_{\parallel}))$. The parameter $k$ characterizes the anisotropy of the particle. For $k=0$ we have the Einstein-de Haas effect when the particle starts to rotate along the short axis as the magnetic field is suddenly turned on.
\begin{figure}
	\centering
	\includegraphics[width=0.9\textwidth]{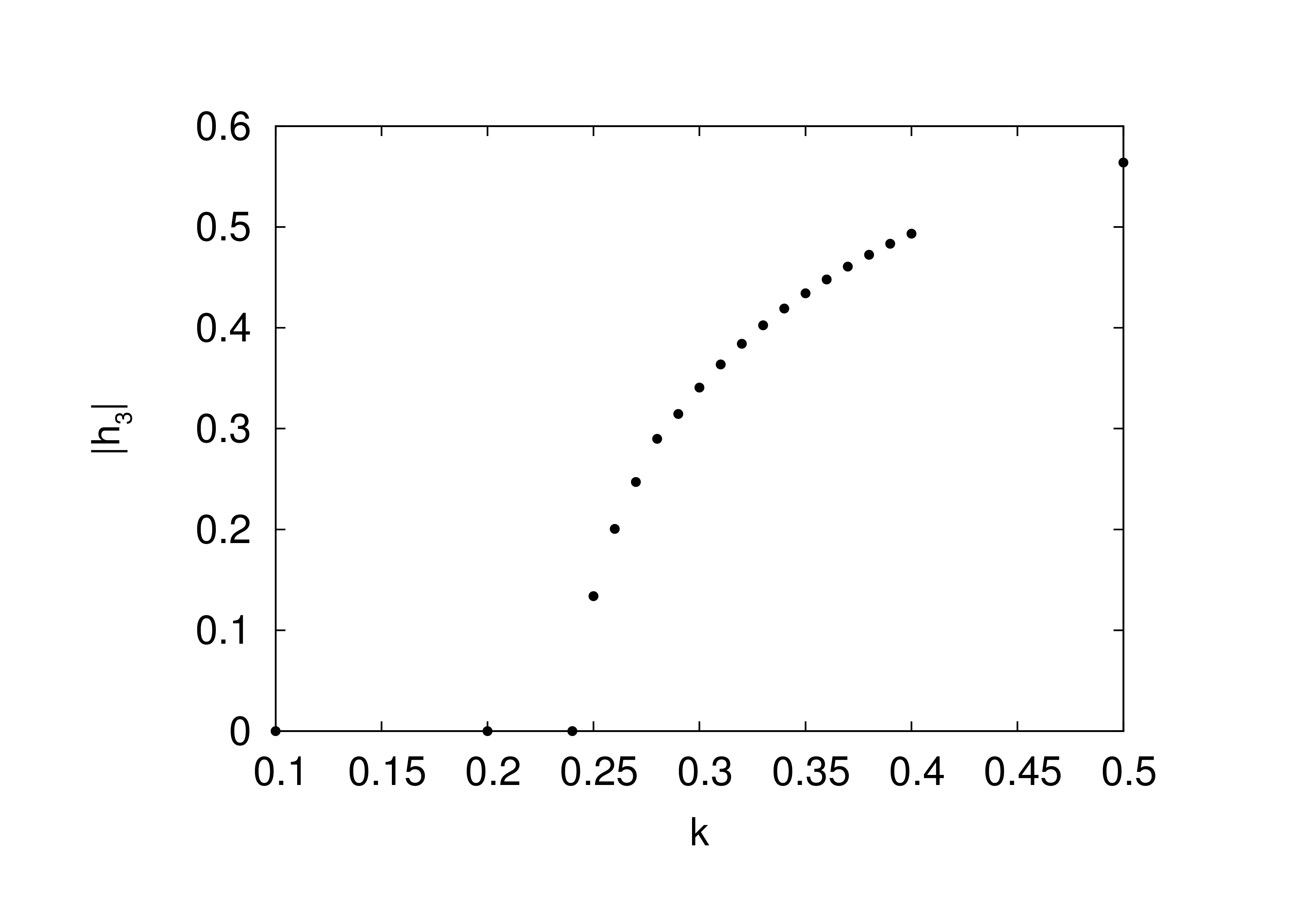}
	\caption{The applied field component along the long axis in steady state, as a function of the anisotropy parameter $k$. $\omega_{0}=0.6;~\sigma=5;~\delta=2;$. }
	\label{fig:7}
\end{figure}
By numerical calculation we show that there is a critical value of $k_{c}$ below which the Einstein-de Haas effect occurs as described above.
This is illustrated by the bifurcation diagram in Fig.\ref{fig:7} ($\omega_{0}=0.6;~\sigma=5;~\delta=2$), which shows that rotation along the short axis is established at $k<k_{c}\simeq 0.24$. At  $k>k_{c}$ a stationary precession regime arises. Introducing $\vec{z}=\vec{h}+\vec{h}_{a}-\vec{\Omega}-\vec{m}$ we see similarly as above that the only solution in the stationary case is $\vec{z}=0$. Thus
\begin{equation}
\frac{d'\vec{L}}{dt}=-\omega_{0}[\vec{\Omega}\times\vec{L}]-\delta\omega_{0}[\vec{m}\times\vec{h}_{a}]
\label{Eq:28}
\end{equation}
It is easy to see that $\Omega_{3}$ and $m_{3}$ are constants of motion in the stationary case. Then we have
\begin{equation}
P_{1}=\frac{m_{1}\delta k}{\sigma-1}\frac{m_{3}}{\Omega_{3}}-\Omega_{1}=0;~P_{2}=\frac{m_{2}\delta k}{\sigma-1}\frac{m_{3}}{\Omega_{3}}-\Omega_{2}=0~.
\label{Eq:29}
\end{equation}
\begin{figure}
	\centering
	\includegraphics[width=0.9\textwidth]{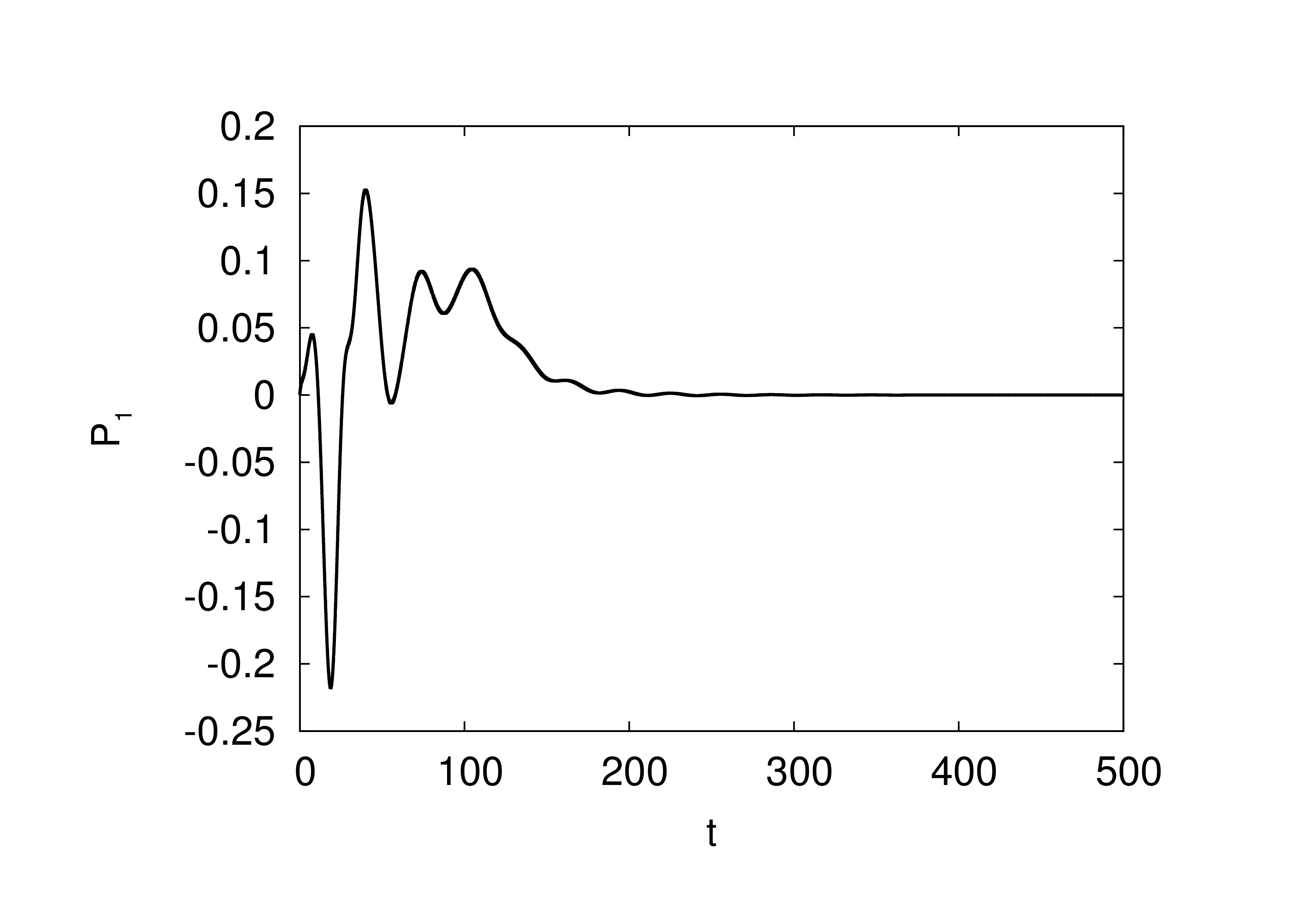}
	\caption{$P_{1}$ as a function of time. $\omega_{0}=0.6;~\sigma=5;~\delta=2;~k=0.4$. }
	\label{fig:8a}
\end{figure}
\begin{figure}
	\centering
	\includegraphics[width=0.9\textwidth]{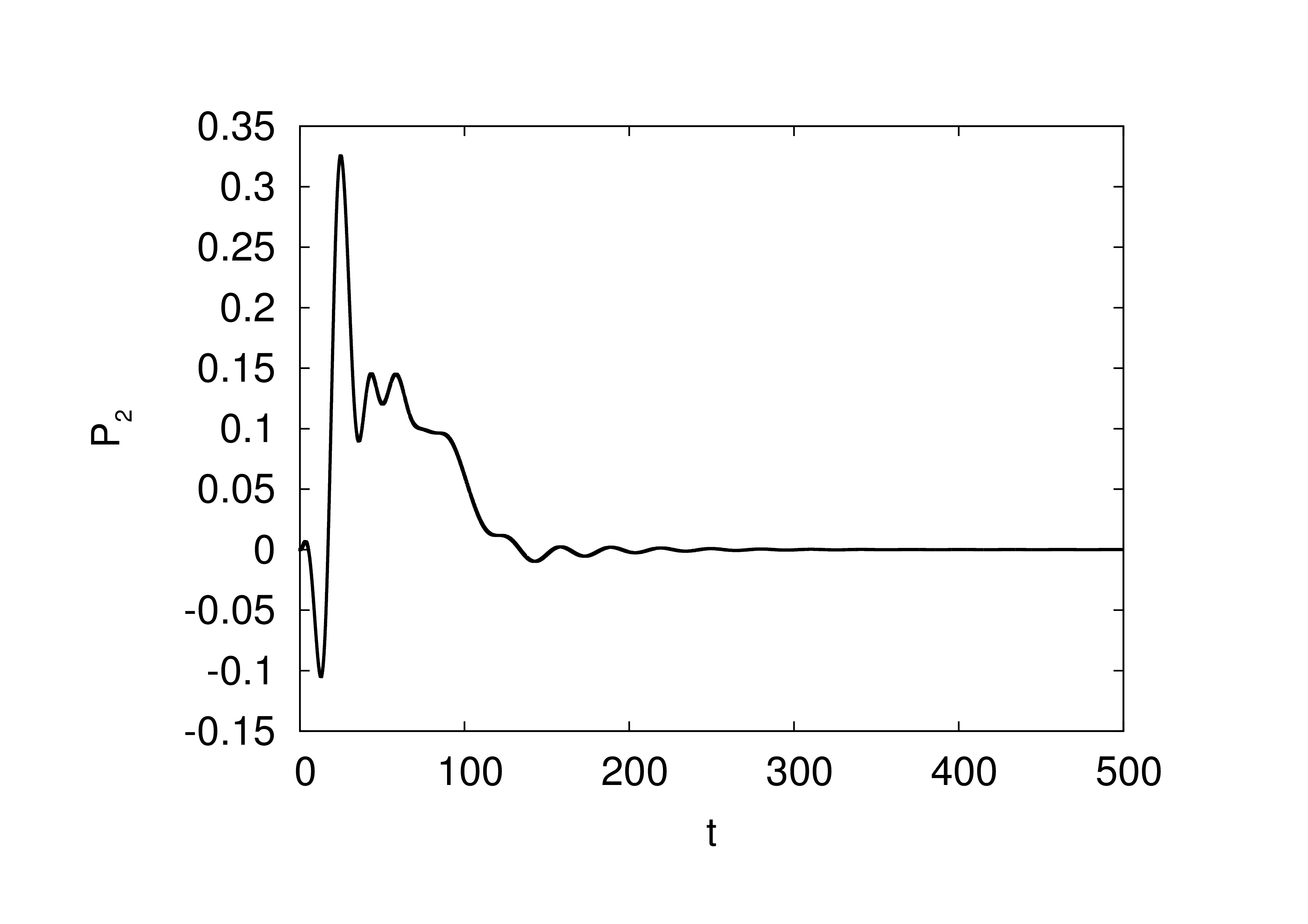}
	\caption{$P_{2}$ as a function of time. $\omega_{0}=0.6;~\sigma=5;~\delta=2;~k=0.4$. }
	\label{fig:8b}
\end{figure}
The left parts of the relations (\ref{Eq:29})$P_{1},P_{2}$ as function of $t$ are shown in Fig.\ref{fig:8a},\ref{fig:8b} $(\omega_{0}=0.6;\delta=2;\sigma=5,k=0.4)$. We see that the relations (\ref{Eq:29}) are satisfied in the stationary case. The angular frequency of the precessional motion may be found from the relation $\Omega_{pr}=\frac{\Omega_{3}}{\cos{(\vartheta)}}$.

\section{Single domain ferromagnetic particle}
\subsection*{Model}
The energy of a single domain spherical ferromagnetic particle reads ($\vec{e}=\vec{m}/m$, the magnetic moment $m$ is considered to be constant, $\vec{n}$ is the unit vector along the magnetic anisotropy axis, $V$ is the volume of the particle and $K$ measures the coupling strength of the magnetic moment with the anisotropy axis)
\begin{equation}
E=-\vec{m}\cdot\vec{H}-\frac{1}{2}KV(\vec{e}\cdot\vec{n})^{2}+\frac{1}{2}I\vec{\Omega}^2~.
\label{Eq:30}
\end{equation}
We derive $dE/dt$ with the condition of total angular momentum balance
\begin{equation}
I\frac{d\vec{\Omega}}{dt}+\frac{1}{\gamma}\frac{d\vec{m}}{dt}=[\vec{m}\times\vec{H}]~.
\label{Eq:31}
\end{equation}
and the kinematic condition
\begin{equation}
\frac{d\vec{n}}{dt}=[\vec{\Omega}\times\vec{n}]~.
\label{Eq:32}
\end{equation}
As a result we have
\begin{equation}
\frac{dE}{dt}=-\Bigl(\vec{H}+\vec{H}_{a}+\frac{1}{\gamma}\vec{\Omega}\Bigr)\Bigl(\frac{d\vec{m}}{dt}-[\vec{\Omega}\times\vec{m}]\Bigr)~.
\label{Eq:33}
\end{equation}
Here $\vec{H}_{a}=\frac{KV}{m}(\vec{e}\cdot\vec{n})\vec{n}$ is the effective field due to the magnetic anisotropy.

In the dissipationless case
\begin{equation}
\frac{d\vec{m}}{dt}=\gamma[\vec{m}\times(\vec{H}+\vec{H}_{a})]
\label{Eq:34}
\end{equation}
and the magnetic moment carries out Larmor precession around the effective field $\vec{H}+\vec{H}_{a}$. 
\textbf{In the presence of dissipation}
\begin{equation}
\Bigl(\frac{d\vec{m}}{dt}\Bigr)_{diss}=[\vec{\Omega}_{d}\times\vec{m}]~.
\label{Eq:35}
\end{equation}
As a result for the entropy production we have
\begin{equation}
T\frac{dS}{dt}=-\frac{dE}{dt}=\vec{\Omega}_{d}\cdot[\vec{m}\times\Bigl(\vec{H}+\vec{H}_{a}+\frac{1}{\gamma}\vec{\Omega}\Bigr)]~.
\label{Eq:36}
\end{equation}
Using the linear phenomenological law for coupling the thermodynamic force $[\vec{m}\times\Bigl(\vec{H}+\vec{H}_{a}+\frac{1}{\gamma}\vec{\Omega}\Bigr)]$ and flux $\vec{\Omega}_{d}$  we postulate
\begin{equation}
\vec{\Omega}_{d}=\frac{1}{\zeta}[\vec{m}\times\Bigl(\vec{H}+\vec{H}_{a}+\frac{\vec{\Omega}}{\gamma}\Bigr)]~.
\label{Eq:37}
\end{equation}
As a result we obtain a set of coupled equations for the magnetic moment and the particle
\begin{equation}
\frac{d\vec{m}}{dt}=\gamma[\vec{m}\times(\vec{H}+\vec{H}_{a})]+\frac{1}{\zeta}[[\vec{m}\times(\vec{H}+\vec{H}_{a}+\frac{1}{\gamma}\vec{\Omega})]\times\vec{m}]~,
\label{Eq:38}
\end{equation}
\begin{equation}
I\frac{d\vec{\Omega}}{dt}=-[\vec{m}\times\vec{H}_{a}]-\frac{1}{\zeta\gamma}[[\vec{m}\times(\vec{H}+\vec{H}_{a}+\frac{1}{\gamma}\vec{\Omega})]\times\vec{m}]
\label{Eq:39}
\end{equation}
and
\begin{equation}
\frac{d\vec{n}}{dt}=[\vec{\Omega}\times\vec{n}]~.
\label{Eq:40}
\end{equation}
The set Eqs.(\ref{Eq:38},\ref{Eq:39},\ref{Eq:40})
contains both the Einstein-de Haas (rotation of the particle due to change of internal angular momentum) and Barnett (the appearance in a body frame of reference of an additional field $\vec{\Omega}/\gamma$) effects.

The following scalings are introduced $\tilde{t}=\omega_{E}t;~\omega_{E}=m/(|\gamma I);~\vec{\Omega}=|\gamma|H\tilde{\vec{\Omega}}$. Introducing parameter $\beta$ according to $\frac{1}{\zeta}=\frac{\beta|\gamma}{m}$, denoting $\varepsilon=\frac{|\gamma|H}{\omega_{E}}$ and $\vec{H}=H\vec{h}$ set of equations rewritten in dimensionless form reads
\begin{equation}
\frac{d\vec{e}}{dt}=-\varepsilon[\vec{e}\times\Bigl(\vec{h}+\frac{H_{a}}{H}(\vec{e}\cdot\vec{n})\vec{n}\Bigr)]+\varepsilon\beta[[\vec{e}\times\Bigl(\vec{h}+\frac{H_{a}}{H}(\vec{e}\cdot\vec{n})\vec{n}-\vec{\Omega}\Bigr)]\times\vec{e}]~,
\label{Eq:41}
\end{equation}
\begin{equation}
\frac{d\vec{\Omega}}{dt}=-\frac{H_{a}}{H}(\vec{e}\cdot\vec{n})[\vec{e}\times\vec{n}]+\beta[[\vec{e}\times\Bigl(\vec{h}+\frac{H_{a}}{H}(\vec{e}\cdot\vec{n})\vec{n}-\vec{\Omega}\Bigr)]\times\vec{e}]~,
\label{Eq:42}
\end{equation}
\begin{equation}
\frac{d\vec{n}}{dt}=\varepsilon[\vec{\Omega}\times\vec{n}]~.
\label{Eq:43}
\end{equation}

\subsection*{Dissipationless case}
Equations (\ref{Eq:41},\ref{Eq:42},\ref{Eq:43}) may be considerably simplified in the dissipationless case. If $\vec{\Omega}_{d}=0$ Eqs.(\ref{Eq:41},\ref{Eq:42}) give
\begin{equation}
\frac{d\vec{e}}{dt}=\varepsilon[\vec{\Omega}\times\vec{e}]
\label{Eq:44}
\end{equation}
and
\begin{equation}
\frac{d\vec{\Omega}}{dt}=[\vec{e}\times\vec{h}]+[\vec{\Omega}\times\vec{e}]~.
\label{Eq:45}
\end{equation}
The shortened equations contain two variables $\vec{e},\vec{\Omega}$ instead of $\vec{e},\vec{\Omega},\vec{n}$ of the full model since the absence of dissipation means that orientation of magnetic moment with respect to $\vec{n}$ remains constant (magnetic moment rotates with the particle).

The dynamics of the particle depends on the ratio of the precession frequency to the Einstein frequency $\varepsilon$. Since the set (\ref{Eq:44},\ref{Eq:45}) has two integrals of motion it is possible to solve the equation of motion in quadratures. The first integral is the energy, which in dimensionless variables reads $\varepsilon\vec{\Omega}^{2}/2-\vec{e}\cdot\vec{h}$. It is easy to see that $d(\vec{\Omega}\cdot\vec{e})/dt=0$. Then $\vec{\Omega}\cdot\vec{e}=0$ if we start at initial condition $\vec{\Omega}=0$. Then by derivation with respect to time of $\varepsilon\vec{\Omega}=[\vec{e}\times\frac{d\vec{e}}{dt}]$ and using the law of total angular momentum it is possible to obtain
\begin{equation}
\vec{A}=\Bigl[\vec{e}\times\frac{d^{2}\vec{e}}{dt^{2}}\Bigr]-\frac{d\vec{e}}{dt}-\varepsilon[\vec{e}\times\vec{h}]=0~.
\label{Eq:46}
\end{equation}
$A_{z}=0$  gives
\begin{equation}
\frac{d(\sin^{2}{(\vartheta)}\dot{\varphi}-\cos{(\vartheta)})}{dt}=0~.
\label{Eq:47}
\end{equation}
Calculation of $\cos{(\varphi)}A_{y}-\sin{(\varphi)}A_{x}$ gives
\begin{equation}
\frac{d^{2}\vartheta}{dt^{2}}-\Bigl(\frac{d\varphi}{dt}\Bigr)^2\cos{(\vartheta)}\sin{(\vartheta)}+\varepsilon\sin{(\vartheta)}-\sin{(\vartheta)}\frac{d\varphi}{dt}=0~.
\label{Eq:48}
\end{equation}
The set of equations (\ref{Eq:47},\ref{Eq:48}) may be put in the Lagrangian form taking
$L=\varepsilon(T-V+G)$, where $T$ is the kinetic energy, $V$ is the potential energy, $G$ is the gyroscopic term and
\begin{eqnarray}
T=\frac{1}{2\varepsilon}(\dot{\vartheta}^{2}+\sin^{2}{(\vartheta)}\dot{\varphi}^{2}) \\ \nonumber
V=-\cos{(\vartheta)}\\ \nonumber
G=-\frac{1}{\varepsilon}\dot{\varphi}\cos{(\vartheta)}~.
\label{Eq:49}
\end{eqnarray}
The Lagrange equations
\begin{equation}
\frac{d}{dt}\frac{\partial L}{\partial \dot{\varphi}}-\frac{\partial L}{\partial \varphi}=0;~
\frac{d}{dt}\frac{\partial L}{\partial \dot{\vartheta}}-\frac{\partial L}{\partial \vartheta}=0
\label{Eq:50}
\end{equation}
give (\ref{Eq:47},\ref{Eq:48}) respectively.
The set of equations (\ref{Eq:50}) may be put in Hamiltonian form, which may be useful due to the possibility to apply symplectic algorithms to numerically solve the equations of motion, satisfying energy conservation even in the case of long trajectories.
Taking $p_{\vartheta}=\frac{\partial L}{\partial \dot{\vartheta}}$ and $p_{\varphi}=\frac{\partial L}{\partial \dot{\varphi}}$ for the Hamiltonian we obtain
\begin{equation}
H=p_{\vartheta}\dot{\vartheta}+p_{\varphi}\dot{\varphi}-L=\frac{1}{2}\Bigl(p_{\vartheta}^{2}+\frac{(p_{\varphi}+\cos{(\vartheta)})^{2}}{\sin^{2}{(\vartheta)}}\Bigr)-\varepsilon\cos{(\vartheta)}~.
\label{Eq:51}
\end{equation}
We should note that in this Hamiltonian the momentum and position are not separated, which puts considerable constraints on the potential choices of symplectic algorithm that can be employed to numerically solve this system.

\begin{figure}
	\centering
	\includegraphics[width=0.9\textwidth]{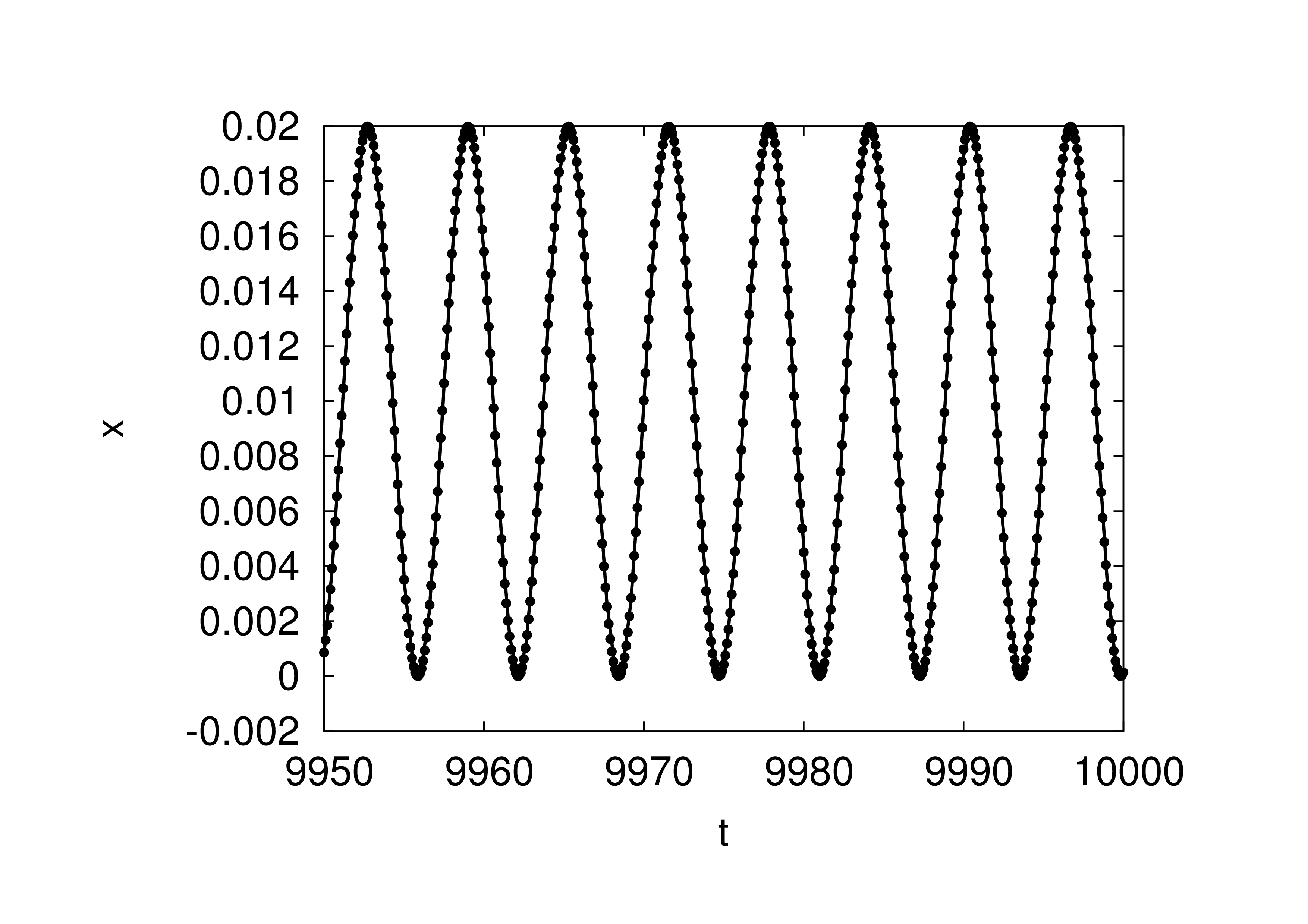}
	\caption{Small amplitude nutation at the precessional motion. $\varepsilon=0.01$. Solid line fit by the solution $x=0.01(1-cos(0.013-1.0003t)$ }
	\label{fig:prec}
\end{figure}

The Hamiltonian form of the equations allows us to consider the precessional motion of the magnetic moment for small $\varepsilon$. In the case when $\vartheta=\pi/2+x$ ($x\ll 1$) we have (up to second order terms)
\begin{equation}
H=\frac{1}{2}\Bigl(p_{\vartheta}^{2}+p_{\varphi}^{2}\Bigr)+(\varepsilon-p_{\varphi})x+\frac{1}{2}(1+p_{\varphi}^{2})x^{2}~,
\label{Eq:52}
\end{equation}
For initial conditions we take $\vec{\Omega}=0;~\vec{e}=(1,0,0)$ $p_{\varphi}=0$ and we have the equation for $x$
\begin{equation}
\frac{d^{2}x}{dt^{2}}+x=-\varepsilon~.
\label{Eq:53}
\end{equation}
Its solution at $x(0);~\dot{x}(0)$ is $x=-\varepsilon(1-\cos{(t)})$.
Small oscillations around $\vartheta=\pi/2$ as found numerically at given initial conditions are shown in Fig.\ref{fig:prec} and are rather closely matched by the solution of Eq.(\ref{Eq:53}).

\begin{figure}
	\centering
	\includegraphics[width=0.9\textwidth]{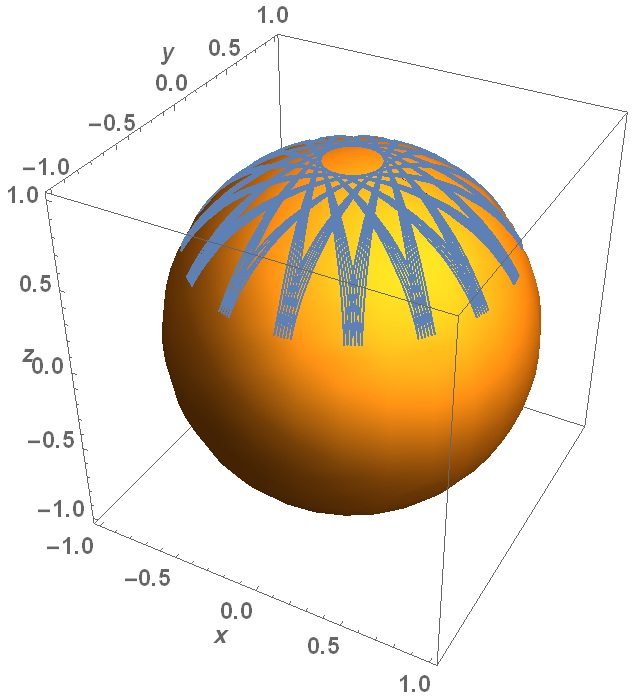}
	\caption{Nutation at precessional motion of the magnetic moment. Initial conditions $\vec{e}(0)=(1/2,1/\sqrt{2},1/2);~\vec{\Omega}(0)=0.$ $\varepsilon=10$.}
	\label{fig:nut}
\end{figure}

\begin{figure}
	\centering
	\includegraphics[width=0.9\textwidth]{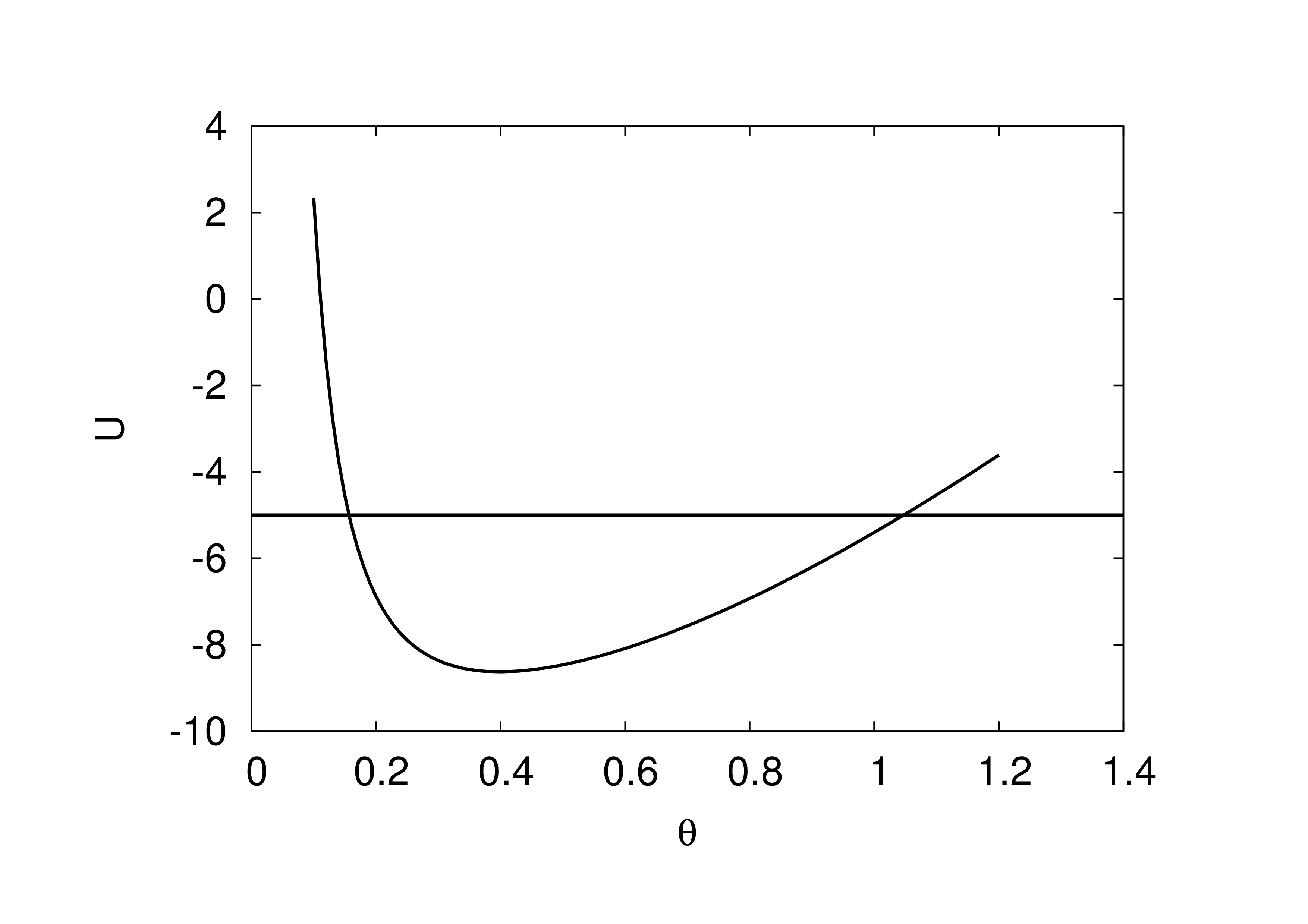}
	\caption{The effective potencial energy as a function of the angle of the magnetic moment with the applied field. Values of the integrals $E,p_{\varphi}$ correspond to the initial conditions $\vec{e}(0)=(1/2,1/\sqrt{2},1/2);~\vec{\Omega}(0)=0.$ Horizontal line shows the value of the energy $E=-5$ corresponding to the initial condition.}
	\label{fig:pot}
\end{figure}

A precession-nutation regime takes place at large $\varepsilon$ values. The integral of motion $H=E$ allows us to integrate the equations of motion. Introducing the effective potential energy
\begin{equation}
U(\vartheta)=\frac{(p_{\varphi}+\cos{(\vartheta)})^{2}}{2\sin^{2}{(\vartheta)}}-\varepsilon\cos{(\vartheta)}~.
\label{Eq:54}
\end{equation}
the integral of motion reads
\begin{equation}
\frac{1}{2}\dot{\vartheta}^{2}+U(\vartheta)=E~.
\label{Eq:55}
\end{equation}
The trajectory $\vec{e}(t)$ on the unit sphere at $\varepsilon=10$ and initial conditions $\vec{e}(0)=(1/2,1/\sqrt{2},1/2)$ and $\vec{\Omega}=0$ is shown in Fig.\ref{fig:nut}. We see that in agreement with the effective potential energy (Fig.\ref{fig:pot}) the nutation angle $\vartheta\in [0.157;\pi/3]$ where the effective potential energy $U(\vartheta)<E=-5$ for given initial conditions when $p_{\varphi}=-1/2$. 

\section{Conclusions}
In summary, we have formulated a set of ordinary differential equations describing the dynamics of magnetic particles taking into account gyromagnetic effects. In the case of an ellipsoidal paramagnetic particle it is found that, despite starting from a quiet non-magnetized state, after the magnetic field is switched on a rotation along the short axis is established. This is confirmed by the stability analysis of the fixed points of the corresponding ODE.
In the case of a ferromagnetic particle we found the integrals of motion in the dissipationless case, which allowed us to integrate the ODE describing the precession and nutation of ferromagnetic particle.

\section{Acknowledgments}
Authors are thankful to D.Budker for fruitful discussions.
This work is funded by QuantERA grant LEMAQUME funded by the QuantERA//ERA-NET Cofund in Quantum Technologies implemented within the European Union's Horizon2020 Programme. Authors acknowledge support by Scientific Council of Latvia Grant No. lzp-2020/1-0149.

\section{Appendix A}
Simulation animations of figures 1 and 14 can be found in the supplementary materials.

Animation 1 illustrates the Einstein-de Haas effect for an ellipsoidal paramagnetic particle by showing the trajectory of the unit vector $\vec{h}$ on the unit sphere over time. The end state of the animation, at which point the long axis has oriented itself perpendicularly to the applied field, corresponds to Fig.1 in the main paper. $\theta=0.1$, $\omega_0=0.6$, $\sigma=5$, $\delta=1.5$.

Animations 2-5 illustrate the dynamics of the precession-nutation regime that is observed in the dissipationless case. First, at $\epsilon=0.1$ only small amplitude nutation takes place, similarly to the case shown in Fig.13 in the main paper. As $epsilon$ is increased, a more complex pattern of larger amplitude nutations becomes apparent (Fig.14). All animations are recorded at the same speed to illustrate the fact that as $\epsilon$ is increased, the oscillations speed up accordingly. Above $\epsilon=100$ the model starts exhibiting signs of numerical instability.

Animation 1 (einstein01.mp4)

Animation 2 (nutation-eps0.1.mp4)

Animation 3 (nutation-eps1.mp4)

Animation 4 (nutation-eps10.mp4)

Animation 5 (nutation-eps100.mp4)

\end{document}